\documentclass[12pt]{article}
\usepackage{amsmath}

\begin{document}

\title{{\vspace{-.8cm}\textbf{\Large Identification of all Hardy-type correlations for\\
\vspace{-.1cm}two photons or particles with spin 1/2\vspace{.1cm}}}}
\author{Jos\'{e} L. Cereceda  \\ 
\textit{C/Alto del Le\'{o}n 8, 4A, 28038 Madrid, Spain}  \\
\small{Electronic address: jl.cereceda@teleline.es}}
\date{\today}
%%%\date{May 5, 2001}

\maketitle
\begin{abstract}
By using an alternative, equivalent form  of the CHSH inequality and making extensive use of the experimentally testable property of physical locality we determine the 64 different Bell-type inequalities (each one involving four joint probabilities) into which Hardy's nonlocality theorem can be cast. This allows one to identify all the two-qubit correlations which can exhibit \mbox{Hardy-type} nonlocality.

\vspace{.3cm}
\noindent \textit{Key words:} EPRB-type experiment, joint probability, physical locality, Bell's inequality, quantum mechanics, Hardy's nonlocality theorem.
\end{abstract}

\section{Introduction and notation}

In 1992, Lucien Hardy [1] developed a proof of nonlocality for two particles without using inequalities, with each of the particles living in an effective two-dimensional Hilbert state space. A more systematic version of this proof was given by Hardy himself [2], and then by Goldstein [3], Jordan [4], and Mermin [5], among other authors. Hardy's nonlocality proof involves an experimental set-up of the Einstein-Podolsky-Rosen-Bohm-type [6,7]: Two correlated particles 1 and 2 fly apart in opposite directions from some common source such that each of the particles subsequently enters its own measuring apparatus (located at wings $A$ and $B$ respectively) which can measure either one of two physical observables at a time---$a_1$ or $a_2$ for particle 1 and $b_1$ or $b_2$ for particle 2. Conventionally, it is supposed that a measurement of each one of these observables gives the possible outcomes either ``$+1$'' or ``$-1$''. Also, a parameter setting $a_j$ ($b_k$), $j,k = 1,2$, for the measuring apparatus at side $A$ ($B$) is meant to signify a measurement of the observable $a_j$ ($b_k$) at that side. So the experiment consists in measuring a large number of particle pairs, with the setting at wing $A$ ($B$) alternated between $a_1$ and $a_2$ ($b_1$ and $b_2$). The basic entity to be considered is the joint probability $p(a_j =m,b_k =n)$ that the outcome of the measurement of $a_j$ on particle 1 is $m$, and that the outcome of the measurement of $b_k$ on the paired particle 2 is $n$, where $j,k = 1,2$, and $m,n=\pm 1$. Each of these probabilities fulfills the property
\begin{equation}
0 \leq  p(a_j =m, b_k =n) \leq 1 .
\end{equation}
On the other hand, throughout this work we shall assume ideal behaviour of the measuring apparatuses and, in particular, perfect efficiency of the detection equipment. This requires the observable probabilities $p(a_j =m,b_k =n)$ to satisfy the normalization condition
\begin{equation}
\sum_{m,n =\pm 1}  p(a_j =m, b_k =n) =1 ,
\end{equation}
for any $j,k = 1 \text{ or } 2$.

Hardy's \textit{gedanken\/} experiment is based on the vanishing of three appropriate joint probabilities $p(a_j =m,b_k =n)$ while a fourth one is nonzero [1-5,8]. In practice, however, this leads to the problem of experimentally measuring a null event. Indeed, \textit{even\/} with perfect detectors, it is not possible in any case to achieve a true ``zero'' value for the various probabilities because the number of runs in a real experiment is necessarily finite [5]. Moreover, for imperfect detectors, inequalities are necessary in order to show that the experimental errors do not wash out the logical contradiction which arises when confronting the predictions entailed by quantum mechanics with those based on the classical notion of local realism [9]. In [5] (see also [10]) Mermin derived an inequality of the Clauser-Horne (CH) type [11-13] into which Hardy's argument can be cast. This kind of inequality has been used, for example, in the experimental demonstration of Hardy's nonlocality reported in Ref.\ 9. (It is to be noted that, however, this and all other actual experiments testing Bell-type inequalities have relied on one or another sort of supplementary assumption (like the fair-sampling assumption) in order to deal with the detection efficiency loophole [14].\,\footnote{%%%
A notable exception to this statement is the recent experiment in Ref.\ 15 performed with trapped ions in which the detection efficiency was high enough for a Bell's inequality to be violated without requiring the assumption of fair sampling.})

In the present paper, by taking a significantly different approach, we derive systematically all the Bell-type inequalities of the kind discovered by Mermin for the Hardy experiment. There turn out to be 64 such inequalities (see Eqs.\ (27)-(30) in Sec.\ 4). Provided with such inequalities, one can determine all the two-qubit correlations which can exhibit Hardy-type nonlocality. However, in contrast to the derivation presented by Mermin (see Appendix B of [5]), our inequalities are obtained directly from the well-known inequalities of Clauser, Horne, Shimony, and Holt (CHSH) [12,13,16] (more precisely, they are derived from the completely equivalent version of the CHSH inequalities presented in Sec.\ 2). Our method also rests crucially on the unambiguous, experimentally testable property of physical locality (see Sec.\ 3 for a statement of the meaning of this property for the EPR-Bohm experiment). Also, in Sec.\ 3, we write down all the relevant constraints imposed by the physical locality condition on the joint probabilities involved in the EPR-Bohm experiment. In Sec.\ 4, such constraints will be used in conjunction with the CHSH inequalities of Sec.\ 2 to obtain all the 64 inequalities mentioned above. The fact that the CH-type inequality derived by Mermin for the Hardy experiment can be extracted directly from the CHSH inequality should not be surprising because, as shown in [10], the CH and CHSH inequalities are equivalent \textit{provided\/} physical locality holds. (A further, more elaborate demonstration of this equivalence is given in Sec.\ 5.) This fact, however, calls in doubt whether it is appropriate to label Mermin's inequality a ``CH-type'' inequality, as opposed to a ``CHSH-type'' inequality, since, as shown in this paper, the Bell-type inequality derived by Mermin for the Hardy experiment is completely equivalent to both the CH and CHSH inequalities, and can actually be obtained from either one of them (provided physical locality holds). In the light of these results, in Sec.\ 6 we examine a seemingly paradoxical situation discussed by Mermin in Ref.\ 10 in the context of a question raised by Popescu and Rohrlich in Ref.\ 17. We finally conclude in Sec.\ 7.

In order to abbreviate the notation the various probabilities $p(a_j =m, b_k =n)$ will henceforth be denoted by the respective shorthands $p1, p2,\ldots,p16$, according to the convention
\begin{align}
p1 &\equiv p(a_1=1, b_1=1),  & p2 &\equiv p(a_1=1, b_1=-1),  \nonumber  \\
p3 &\equiv p(a_1=-1, b_1=1), & p4 &\equiv p(a_1=-1, b_1=-1), \nonumber  \\
p5 &\equiv p(a_1=1, b_2=1),  & p6 &\equiv p(a_1=1, b_2=-1),  \nonumber   \\
p7 &\equiv p(a_1=-1, b_2=1), & p8 &\equiv p(a_1=-1, b_2=-1),  \nonumber  \\[-.3cm]
&  &  &  \\[-.3cm]
p9 &\equiv p(a_2=1, b_1=1),  & p10 &\equiv p(a_2=1, b_1=-1),  \nonumber  \\
p11 &\equiv p(a_2=-1, b_1=1),& p12 &\equiv p(a_2=-1, b_1=-1), \nonumber  \\
p13 &\equiv p(a_2=1, b_2=1), & p14 &\equiv p(a_2=1, b_2=-1),  \nonumber  \\
p15 &\equiv p(a_2=-1, b_2=1),& p16 &\equiv p(a_2=-1, b_2=-1). \nonumber   
\end{align}
We also introduce for later convenience the following eight sums of joint probabilities:
\begin{align}
&\Sigma_1 = p1+p4+p5+p8+p9+p12+p14+p15,  \nonumber  \\
&\Sigma_{1}^{\prime} = p2+p3+p6+p7+p10+p11+p13+p16,  \nonumber  \\
&\Sigma_2 = p1+p4+p5+p8+p10+p11+p13+p16,  \nonumber  \\
&\Sigma_{2}^{\prime} = p2+p3+p6+p7+p9+p12+p14+p15,  \nonumber  \\[-.3cm]
&  \\[-.3cm]
&\Sigma_3 = p1+p4+p6+p7+p9+p12+p13+p16,  \nonumber  \\
&\Sigma_{3}^{\prime} = p2+p3+p5+p8+p10+p11+p14+p15,  \nonumber  \\
&\Sigma_4 = p2+p3+p5+p8+p9+p12+p13+p16,  \nonumber  \\
&\Sigma_{4}^{\prime} = p1+p4+p6+p7+p10+p11+p14+p15.  \nonumber
\end{align}

\section{An alternative form of the CHSH inequality}

The standard form of the CHSH inequality reads [12,13,16]
\begin{equation}
-2 \leq \Delta_i \leq 2  ,
\end{equation}
where the quantity $\Delta_i$ can be any one of the following four sums of correlations
\begin{align}
& \Delta_1 = c(a_1,b_1) + c(a_1,b_2) + c(a_2,b_1) - c(a_2,b_2), \nonumber \\
& \Delta_2 = c(a_1,b_1) + c(a_1,b_2) - c(a_2,b_1) + c(a_2,b_2),  \nonumber \\[-.3cm]
& \\[-.3cm]
& \Delta_3 = c(a_1,b_1) - c(a_1,b_2) + c(a_2,b_1) + c(a_2,b_2),  \nonumber \\
& \Delta_4 = -c(a_1,b_1) + c(a_1,b_2) + c(a_2,b_1) + c(a_2,b_2),  \nonumber
\end{align}
the correlation coefficient $c(a_j,b_k)$ being defined by
\begin{align}
c(a_j ,b_k) = \;\, &p(a_j =1,b_k=1) + p(a_j=-1,b_k=-1)  \nonumber  \\
&-p(a_j=1,b_k=-1) - p(a_j=-1,b_k=1).
\end{align}
Next we present another, equivalent form of the CHSH inequality (5) that will prove useful in Sec.\ 4, where we derive all the relevant Bell-type inequalities which can be used to experimentally test Hardy-type nonlocality. So, substituting Eq.\ (7) in the above expression for $\Delta_i$ and taking into account the normalization condition in Eq.\ (2), it is straightforward to see that the quantity $\Delta_i$ can equivalently be written as
\begin{equation}
\Delta_i = 2(\Sigma_i -2) = 2(2-\Sigma_{i}^{\prime}), \quad i=1,2,3,4\, ,
\end{equation}
where $\Sigma_i$ and $\Sigma_{i}^{\prime}$ are the sums of probabilities defined in (4). Now, from the standard inequality (5) and the expression for $\Delta_i$ in Eq.\ (8), one can readily deduce the following alternative, equivalent form of the CHSH inequality
\begin{equation}
1 \leq \Sigma_i, \Sigma_{i}^{\prime} \leq 3 , \quad i=1,2,3,4\, .
\end{equation}
The maximum (minimum) value of either $\Sigma_i$ or $\Sigma_{i}^{\prime}$ predicted by quantum mechanics is $2+\sqrt{2}$ [$2-\sqrt{2}$] which occurs when each of the terms in either $\Sigma_i$ or $\Sigma_{i}^{\prime}$ attains the value $(2+\sqrt{2})/8$ [$(2-\sqrt{2})/8$]. Please note, however, that whenever $\Sigma_i$ reaches its maximum quantum mechanical value $2+\sqrt{2}$ then $\Sigma_{i}^{\prime}$ reaches its minimum one $2-\sqrt{2}$, and vice versa, so that the sum $\Sigma_i + \Sigma_{i}^{\prime}$ always amounts to 4 according to the normalization condition (2).

\newpage
\section{Constraints imposed by the experimentally\\ testable property of physical locality}

In addition to the requirements of nonnegativity and normalization the joint probabilities $p1,p2,\ldots,p16$ should satisfy a consistency condition that, following Mermin [10], will be called physical locality (ocassionally, following Hillery and Yurke [18], this latter condition will also be referred to as the requirement of causal communication). For the typical EPR-Bohm experiment we are considering, physical locality requires the statistical distribution of results obtained at wing $A$ ($B$) to be independent of whatever instrumental parameters concerning the other, spatially separated wing $B$ ($A$). In particular, the distribution of results at either of the wings should be independent of the measurement setting of the other. In terms of the probabilities $p(a_j=m,b_k=n)$ this means [18]
\begin{align}
&p(a_j =m) = \sum_{n=\pm 1} p(a_j =m, b_1 =n) = \sum_{n=\pm 1} p(a_j =m, b_2 =n), \tag{10a} \\
&p(b_k =n) = \sum_{m=\pm 1} p(a_1 =m, b_k =n) = \sum_{m=\pm 1} p(a_2 =m, b_k =n), \tag{10b}
\setcounter{equation}{10}
\end{align}
for any $j,k = 1,2$ and $m,n=\pm1$. Condition (10a) states that the probability of obtaining $a_j =m$ is independent of which measurement ($b_1$ or $b_2$) is performed at side $B$. Similarly, condition (10b) states that the probability for $b_k =n$ is independent of which measurement ($a_1$ or $a_2$) is performed at side $A$. A violation of either condition (10a) or (10b) would, in principle, allow the two parties involved in the EPR-Bohm experiment to communicate superluminally since then one of them could determine the observable being measured by the other merely by testing a sufficiently large number of particles at his own measuring station, thus obtaining a bit of classical information from his partner in a spacelike-separated fashion.

A simple demonstration of the fulfillment of relations (10a)-(10b) by the formalism of quantum mechanics is given in the Appendix of Ref.\ 19, and more general demonstrations can be found, for example, in Refs.\ 20 and 21. From the experimental side, there have been a number of recent experiments (see, specifically, Refs.\ 22 and 23) that, directly or indirectly, confirm the validity of the property of physical locality in various relativistic configurations.

Now we can write down explicitly the constraints imposed by the requirements of normalization (cf.\ Eq.\ (2)) and causal communication (cf.\ Eqs.\ (10a)-(10b)) as follows [19]
\begin{align}
& p1+p2+p3+p4 =1,   \nonumber  \\
& p5+p6+p7+p8 =1,    \nonumber  \\
& p9+p10+p11+p12 =1,  \nonumber  \\
& p13+p14+p15+p16 =1,  \nonumber  \\
& p1+p2-p5-p6 =0,      \nonumber \\
& p3+p4-p7-p8 =0,      \nonumber  \\[-.3cm]
&  \\[-.3cm]
& p9+p10-p13-p14 =0,   \nonumber \\
& p11+p12-p15-p16 =0,   \nonumber  \\
& p1+p3-p9-p11 =0,   \nonumber \\
& p2+p4-p10-p12 =0,   \nonumber  \\
& p5+p7-p13-p15 =0,   \nonumber \\
& p6+p8-p14-p16 =0.  \nonumber
\end{align}
The twelve linear equations in (11) determine eight probabilities at most among $p1,p2,\ldots,p16$. So, for example, we can solve the system of equations (11) with respect to the set of variables $\{p2,p3,p6,p7,p10,p11,p13,p16\}$ to get
\begin{align}
p2 &= (1-p1-p4+p5-p8-p9+p12+p14-p15)/2,  \nonumber  \\
p3 &= (1-p1-p4-p5+p8+p9-p12-p14+p15)/2,  \nonumber  \\
p6 &= (1+p1-p4-p5-p8-p9+p12+p14-p15)/2,  \nonumber  \\
p7 &= (1-p1+p4-p5-p8+p9-p12-p14+p15)/2,   \nonumber  \\[-.3cm]
&  \\[-.3cm]
p10 &= (1-p1+p4+p5-p8-p9-p12+p14-p15)/2,  \nonumber  \\
p11 &= (1+p1-p4-p5+p8-p9-p12-p14+p15)/2,  \nonumber  \\
p13 &= (1-p1+p4+p5-p8+p9-p12-p14-p15)/2,  \nonumber  \\
p16 &= (1+p1-p4-p5+p8-p9+p12-p14-p15)/2,  \nonumber  
\end{align}
with inverse relations
\begin{align}
p1 &= (1-p2-p3+p6-p7-p10+p11+p13-p16)/2,  \nonumber  \\
p4 &= (1-p2-p3-p6+p7+p10-p11-p13+p16)/2,  \nonumber  \\
p5 &= (1+p2-p3-p6-p7-p10+p11+p13-p16)/2,  \nonumber  \\
p8 &= (1-p2+p3-p6-p7+p10-p11-p13+p16)/2,  \nonumber  \\[-.3cm]
&  \\[-.3cm]
p9 &= (1-p2+p3+p6-p7-p10-p11+p13-p16)/2,  \nonumber  \\
p12 &= (1+p2-p3-p6+p7-p10-p11-p13+p16)/2,  \nonumber  \\
p14 &= (1-p2+p3+p6-p7+p10-p11-p13-p16)/2,  \nonumber  \\
p15 &= (1+p2-p3-p6+p7-p10+p11-p13-p16)/2.  \nonumber  
\end{align}
Alternatively, we can solve (11) with respect to the set of variables $\{p2,p3,p6,p7,p9,\linebreak p12, p14,p15\}$ to get
\begin{align}
p2 &= (1-p1-p4+p5-p8+p10-p11-p13+p16)/2,  \nonumber  \\
p3 &= (1-p1-p4-p5+p8-p10+p11+p13-p16)/2,  \nonumber  \\
p6 &= (1+p1-p4-p5-p8+p10-p11-p13+p16)/2,  \nonumber  \\
p7 &= (1-p1+p4-p5-p8-p10+p11+p13-p16)/2,  \nonumber  \\[-.3cm]
&  \\[-.3cm]
p9 &= (1+p1-p4-p5+p8-p10-p11+p13-p16)/2,  \nonumber  \\
p12 &= (1-p1+p4+p5-p8-p10-p11-p13+p16)/2,  \nonumber  \\
p14 &= (1+p1-p4-p5+p8+p10-p11-p13-p16)/2,  \nonumber  \\
p15 &= (1-p1+p4+p5-p8-p10+p11-p13-p16)/2,  \nonumber  
\end{align}
with inverse relations
\begin{align}
p1 &= (1-p2-p3+p6-p7+p9-p12-p14+p15)/2,  \nonumber  \\
p4 &= (1-p2-p3-p6+p7-p9+p12+p14-p15)/2,  \nonumber  \\
p5 &= (1+p2-p3-p6-p7+p9-p12-p14+p15)/2,  \nonumber  \\
p8 &= (1-p2+p3-p6-p7-p9+p12+p14-p15)/2,  \nonumber  \\[-.3cm]
&  \\[-.3cm]
p10 &= (1+p2-p3-p6+p7-p9-p12+p14-p15)/2,  \nonumber  \\
p11 &= (1-p2+p3+p6-p7-p9-p12-p14+p15)/2,  \nonumber  \\
p13 &= (1+p2-p3-p6+p7+p9-p12-p14-p15)/2,  \nonumber  \\
p16 &= (1-p2+p3+p6-p7-p9+p12-p14-p15)/2.  \nonumber  
\end{align}
Or else with respect to $\{p2,p3,p5,p8,p10,p11,p14,p15\}$ to get
\begin{align}
p2 &= (1-p1-p4+p6-p7-p9+p12+p13-p16)/2,  \nonumber  \\
p3 &= (1-p1-p4-p6+p7+p9-p12-p13+p16)/2,  \nonumber  \\
p5 &= (1+p1-p4-p6-p7-p9+p12+p13-p16)/2,  \nonumber  \\
p8 &= (1-p1+p4-p6-p7+p9-p12-p13+p16)/2,  \nonumber  \\[-.3cm]
&  \\[-.3cm]
p10 &= (1-p1+p4+p6-p7-p9-p12+p13-p16)/2,  \nonumber  \\
p11 &= (1+p1-p4-p6+p7-p9-p12-p13+p16)/2,  \nonumber  \\
p14 &= (1-p1+p4+p6-p7+p9-p12-p13-p16)/2,  \nonumber  \\
p15 &= (1+p1-p4-p6+p7-p9+p12-p13-p16)/2,  \nonumber  
\end{align}
with inverse relations
\begin{align}
p1 &= (1-p2-p3+p5-p8-p10+p11+p14-p15)/2,  \nonumber  \\
p4 &= (1-p2-p3-p5+p8+p10-p11-p14+p15)/2,  \nonumber  \\
p6 &= (1+p2-p3-p5-p8-p10+p11+p14-p15)/2,  \nonumber  \\
p7 &= (1-p2+p3-p5-p8+p10-p11-p14+p15)/2,  \nonumber  \\[-.3cm]
&  \\[-.3cm]
p9 &= (1-p2+p3+p5-p8-p10-p11+p14-p15)/2,  \nonumber  \\
p12 &= (1+p2-p3-p5+p8-p10-p11-p14+p15)/2,  \nonumber  \\
p13 &= (1-p2+p3+p5-p8+p10-p11-p14-p15)/2,  \nonumber  \\
p16 &= (1+p2-p3-p5+p8-p10+p11-p14-p15)/2.  \nonumber  
\end{align}
Or with respect to $\{p1,p4,p6,p7,p10,p11,p14,p15\}$ to get
\begin{align}
p1 &= (1-p2-p3+p5-p8+p9-p12-p13+p16)/2,  \nonumber  \\
p4 &= (1-p2-p3-p5+p8-p9+p12+p13-p16)/2,  \nonumber  \\
p6 &= (1+p2-p3-p5-p8+p9-p12-p13+p16)/2,  \nonumber  \\
p7 &= (1-p2+p3-p5-p8-p9+p12+p13-p16)/2,   \nonumber  \\[-.3cm]
&  \\[-.3cm]
p10 &= (1+p2-p3-p5+p8-p9-p12+p13-p16)/2,  \nonumber  \\
p11 &= (1-p2+p3+p5-p8-p9-p12-p13+p16)/2,  \nonumber  \\
p14 &= (1+p2-p3-p5+p8+p9-p12-p13-p16)/2,  \nonumber  \\
p15 &= (1-p2+p3+p5-p8-p9+p12-p13-p16)/2,  \nonumber  
\end{align}
with inverse relations
\begin{align}
p2 &= (1-p1-p4+p6-p7+p10-p11-p14+p15)/2,  \nonumber  \\
p3 &= (1-p1-p4-p6+p7-p10+p11+p14-p15)/2,  \nonumber  \\
p5 &= (1+p1-p4-p6-p7+p10-p11-p14+p15)/2,  \nonumber  \\
p8 &= (1-p1+p4-p6-p7-p10+p11+p14-p15)/2,  \nonumber  \\[-.3cm]
&  \\[-.3cm]
p9 &= (1+p1-p4-p6+p7-p10-p11+p14-p15)/2,  \nonumber  \\
p12 &= (1-p1+p4+p6-p7-p10-p11-p14+p15)/2,  \nonumber  \\
p13 &= (1+p1-p4-p6+p7+p10-p11-p14-p15)/2,  \nonumber  \\
p16 &= (1-p1+p4+p6-p7-p10+p11-p14-p15)/2.  \nonumber  
\end{align}

We note the important fact according to which there indeed exist probability distributions $\{p1,p2,\ldots,p16\}$ satisfying \textit{all\/} the constraints in Eqs.\ (12)-(19) required by physical locality and which, at the same time, provide the upper general probabilistic limit for the CHSH sum of correlations, namely $|\Delta_i| =4$. This is achieved if, and only if, each of the eight probabilities entering the sum $\Sigma_i$ ($\Sigma_{i}^{\prime}$) is equal to either $1/2$ or $0$ (correspondingly $0$ or $1/2$). An example of such a distribution is
\begin{equation}
p1=p4=p5=p8=p9=p12=p14=p15=1/2,
\end{equation}
and all other probabilities zero. This example was first pointed out by Popescu and Rohrlich [17] (see also Refs.\ 24 and 25) and we will return to it in Sec.\ 6.

Furthermore it should be mentioned that, in addition to the 64 relationships between joint probabilities in Eqs.\ (12)-(19) which arise as a direct consequence of the fulfillment of the conditions of normalization and causal communication, the condition of nonnegativity in Eq.\ (1) implies a lot of further \textit{nontrivial\/} constraints on its own. So, for example, the nonnegativity of $p13$ in (12) requires that
\begin{equation}
1+p4+p5+p9 \geq p1+p8+p12+p14+p15, 
\end{equation}
where, according to (1), we assume that all the probabilities in (21) are nonnegative. From relations (12)-(19) we can construct 64 inequalities of the form (21). Many more constraints can be established by demanding the nonnegativity of the sum of any combination of probabilities. For example, the nonnegativity of $p10+p16$ requires that (see Eq.\ (12))
\begin{equation}
p9+p15 \leq 1.
\end{equation}
Analogously, the nonnegativity of the sum $p2+p3+p6+p7+p10+p11+p13+p16$ requires that
\begin{equation}
p1+p4+p5+p8+p9+p12+p14+p15 \leq 4,
\end{equation}
and so on. On the other hand, from the nonnegativity of $p3$, $p6$, $p7$, $p10$, and $p16$, and the third relation in (13), we quickly obtain the following inequality
\begin{equation}
2\,p5 -1 \leq p2+p11+p13 .
\end{equation}
Similarly, from Eqs.\ (12)-(19), we can derive a total of 64 inequalities of the form (24). Inequalities of this form are important for what follows because, as we shall see in the next Section, the four probabilities involved in each of them can be used to construct an argument of nonlocality of the kind invented by Hardy. The inequality (24) (more precisely, one inequality of the form (24)) was derived by Mermin in Ref.\ 10, and will be used in Sec.\ 6 where we discuss an important point made by this author [10], namely that the CH and CHSH inequalities need not be equivalent (in fact, they are not) in the absence of physical locality.

We conclude this section by recalling the previously mentioned fact that quantum mechanics satisfies the causal communication constraint expressed by Eqs.\ (10a)-(10b). Consequently, the quantum mechanical predictions for the probabilities $p1,p2,\ldots,p16$ should satisfy in any case the relations in Eqs.\ (12)-(19) as well as all the inequalities induced by the conditions of physical locality and nonnegativity, like that in Eq.\ (24).

\section{Derivation of all Bell-type inequalities for the Hardy experiment}

The 64 relations in Eqs.\ (12)-(19) allow us to express each one of the $\Sigma_i$'s and $\Sigma_{i}^{\prime}$'s, $i=1,2,3,4$, into eight equivalent forms. So, for example, from relations (12) we can deduce at once that
\begin{align}
\Sigma_1  = \,\, & 1+2(p5+p12+p14-p2) = 1+2(p8+p9+p15-p3),  \nonumber   \\
= \,\, & 1+2(p1+p12+p14-p6) = 1+2(p4+p9+p15-p7),   \nonumber  \\[-.3cm]
&  \\[-.3cm]
= \,\, & 1+2(p4+p5+p14-p10) = 1+2(p1+p8+p15-p11),  \nonumber   \\
= \,\, & 1+2(p4+p5+p9-p13) = 1+2(p1+p8+p12-p16).  \nonumber   
\end{align}
Analogously, from relations (13) we have
\begin{align}
\Sigma_1^{\prime}  = \,\, & 1+2(p6+p11+p13-p1) = 1+2(p7+p10+p16-p4),  \nonumber   \\
= \,\, & 1+2(p2+p11+p13-p5) = 1+2(p3+p10+p16-p8),   \nonumber  \\[-.3cm]
&  \\[-.3cm]
= \,\, & 1+2(p3+p6+p13-p9) = 1+2(p2+p7+p16-p12),  \nonumber   \\
= \,\, & 1+2(p3+p6+p10-p14) = 1+2(p2+p7+p11-p15),  \nonumber   
\end{align}
and so on. Thus, from the CHSH inequality in Eq.\ (9), $1 \leq \Sigma_1, \Sigma_{1}^{\prime} \leq 3$, and the above expressions for $\Sigma_1$ and $\Sigma_{1}^{\prime}$ in Eqs.\ (25) and (26), we obtain the following sixteen Bell-type inequalities, each of them involving four joint probabilities
\begin{align}
p1 & \leq   p6+p11+p13 \leq 1+p1,  \nonumber  \\
p2 & \leq   p5+p12+p14 \leq 1+p2,  \nonumber  \\
p3 & \leq  p8+p9+p15 \leq 1+p3,  \nonumber   \\
p4 & \leq  p7+p10+p16 \leq 1+p4,  \nonumber  \\
p5 & \leq  p2+p11+p13 \leq 1+p5,  \nonumber  \\
p6 & \leq  p1+p12+p14 \leq 1+p6,  \nonumber  \\
p7 & \leq  p4+p9+p15 \leq 1+p7,  \nonumber  \\
p8 & \leq  p3+p10+p16 \leq 1+p8,  \nonumber \\[-.3cm]
&  \\[-.3cm]
p9 & \leq  p3+p6+p13 \leq 1+p9,  \nonumber \\
p10 & \leq  p4+p5+p14 \leq 1+p10,  \nonumber \\
p11 & \leq  p1+p8+p15 \leq 1+p11,  \nonumber \\
p12 & \leq  p2+p7+p16 \leq 1+p12,  \nonumber \\
p13 & \leq  p4+p5+p9 \leq 1+p13,  \nonumber \\
p14 & \leq  p3+p6+p10 \leq 1+p14,  \nonumber \\
p15 & \leq  p2+p7+p11 \leq 1+p15,  \nonumber \\
p16 & \leq  p1+p8+p12 \leq 1+p16.  \nonumber
\end{align}

Let us note at this point the following simple rule to construct the 64 Bell-type inequalities from the relations in Eqs.\ (12)-(19): For any given relation the associated inequality $p\text{j} \leq p\text{k} +p\text{l} +p\text{m} \leq 1+p\text{j}$ is formed by taking $p\text{j}$ to be the probability appearing on the left-hand side of the given relation, and by taking $p\text{k}$, $p\text{l}$, and $p\text{m}$ to be the three probabilities on the right-hand side with positive sign. For the sake of completeness, and for convenience for what follows, next we write down the remaining 48 Bell-type inequalities obtained from the basic CHSH-type inequality (9) and the relations (14)-(19). So, by applying the above rule to relations (14) and (15), we obtain
\begin{align}
p1 & \leq   p6+p9+p15 \leq 1+p1,  \nonumber \\
p2 & \leq   p5+p10+p16 \leq 1+p2,  \nonumber \\
p3 & \leq  p8+p11+p13 \leq 1+p3,  \nonumber \\
p4 & \leq  p7+p12+p14 \leq 1+p4,  \nonumber  \\
p5 & \leq  p2+p9+p15 \leq 1+p5,  \nonumber \\
p6 & \leq  p1+p10+p16 \leq 1+p6,  \nonumber \\
p7 & \leq  p4+p11+p13 \leq 1+p7,  \nonumber  \\
p8 & \leq  p3+p12+p14 \leq 1+p8,  \nonumber \\[-.3cm]
&  \\[-.3cm]
p9 & \leq  p1+p8+p13 \leq 1+p9,  \nonumber \\
p10 & \leq  p2+p7+p14 \leq 1+p10,  \nonumber \\
p11 & \leq  p3+p6+p15 \leq 1+p11,  \nonumber \\
p12 & \leq  p4+p5+p16 \leq 1+p12,  \nonumber \\
p13 & \leq  p2+p7+p9 \leq 1+p13,  \nonumber \\
p14 & \leq  p1+p8+p10 \leq 1+p14,  \nonumber \\
p15 & \leq  p4+p5+p11 \leq 1+p15, \nonumber \\
p16 & \leq  p3+p6+p12 \leq 1+p16.  \nonumber
\end{align}
Similarly, from relations (16) and (17) we obtain
\begin{align}
p1 & \leq   p5+p11+p14 \leq 1+p1,  \nonumber \\
p2 & \leq   p6+p12+p13 \leq 1+p2,  \nonumber \\
p3 & \leq  p7+p9+p16 \leq 1+p3,  \nonumber \\
p4 & \leq  p8+p10+p15 \leq 1+p4,  \nonumber \\
p5 & \leq  p1+p12+p13 \leq 1+p5,  \nonumber \\
p6 & \leq  p2+p11+p14 \leq 1+p6,  \nonumber \\
p7 & \leq  p3+p10+p15 \leq 1+p7,  \nonumber  \\
p8 & \leq  p4+p9+p16 \leq 1+p8,  \nonumber \\[-.3cm]
&  \\[-.3cm]
p9 & \leq  p3+p5+p14 \leq 1+p9,  \nonumber \\
p10 & \leq  p4+p6+p13 \leq 1+p10,  \nonumber \\
p11 & \leq  p1+p7+p16 \leq 1+p11,  \nonumber \\
p12 & \leq  p2+p8+p15 \leq 1+p12,  \nonumber \\
p13 & \leq  p3+p5+p10 \leq 1+p13,  \nonumber \\
p14 & \leq  p4+p6+p9 \leq 1+p14,  \nonumber \\
p15 & \leq  p1+p7+p12 \leq 1+p15,  \nonumber \\
p16 & \leq  p2+p8+p11 \leq 1+p16.  \nonumber
\end{align}
Finally, from relations (18) and (19) we obtain
\begin{align}
p1 & \leq   p5+p9+p16 \leq 1+p1,  \nonumber \\
p2 & \leq   p6+p10+p15 \leq 1+p2,  \nonumber \\
p3 & \leq  p7+p11+p14 \leq 1+p3,  \nonumber \\
p4 & \leq  p8+p12+p13 \leq 1+p4,  \nonumber \\
p5 & \leq  p1+p10+p15 \leq 1+p5,  \nonumber \\
p6 & \leq  p2+p9+p16 \leq 1+p6,  \nonumber \\
p7 & \leq  p3+p12+p13 \leq 1+p7,  \nonumber \\
p8 & \leq  p4+p11+p14 \leq 1+p8,  \nonumber \\[-.3cm]
&  \\[-.3cm]
p9 & \leq  p1+p7+p14 \leq 1+p9,  \nonumber \\
p10 & \leq  p2+p8+p13 \leq 1+p10,  \nonumber \\
p11 & \leq  p3+p5+p16 \leq 1+p11,  \nonumber \\
p12 & \leq  p4+p6+p15 \leq 1+p12,  \nonumber \\
p13 & \leq  p1+p7+p10 \leq 1+p13,  \nonumber \\
p14 & \leq  p2+p8+p9 \leq 1+p14,  \nonumber \\
p15 & \leq  p3+p5+p12 \leq 1+p15,  \nonumber \\
p16 & \leq  p4+p6+p11 \leq 1+p16.   \nonumber
\end{align}
Note that there are \textit{four\/} inequalities of the form $p\text{j} \leq p\text{k} +p\text{l} +p\text{m} \leq 1+p\text{j}$ for each $\text{j}=1,2,\ldots,16$.

Hardy's \textit{gedanken\/} experiment applies to the case in which the three intermediate probabilities $p\text{k}$, $p\text{l}$, and $p\text{m}$ are identically equal to zero while the probability $p\text{j}$ is not. This, of course, implies a violation of the inequality $p\text{j} \leq p\text{k} +p\text{l} +p\text{m}$. Therefore the inequalities which are relevant in order to exhibit Hardy-type nonlocality are those involving the lower bound in Eqs.\ (27)-(30). Let us consider for concreteness the following theoretical predictions which are allowed by quantum mechanics 
\begin{align}
p4 &\equiv p(a_1=-1, b_1=-1) =0,  \tag{31a} \\
p5 &\equiv p(a_1=+1, b_2=+1)=0,   \tag{31b} \\
p9 &\equiv p(a_2=+1, b_1=+1)=0,   \tag{31c} \\
p13 &\equiv p(a_2=+1, b_2=+1)>0.  \tag{31d}
\setcounter{equation}{31}
\end{align}
In brief, Hardy's argument for nonlocality proceeds as follows [1-5]: Consider a particular run of
the experiment for which the observables $a_2$ and $b_2$ are measured on particles 1 and 2, respectively, and the results $a_2=1$ and $b_2=1$ are obtained. From Eq.\ (31d), there is a nonzero probability for these joint measurement results to occur. (We further assume that, in any joint measurement, the overall act of measurement conducted at either side is spacelike separated from that conducted at the other.) Thus, by invoking the notion of local realism, and taking into account the constraint in Eq.\ (31c), we can deduce that a result $b_1=-1$ would have been obtained if, instead of $b_2$, the observable $b_1$ had been measured on particle 2. Likewise, from Eq.\ (31b), and according to local realism, we can assert that a result $a_1=-1$ would have been obtained in a measurement of $a_1$ on particle 1. In this way, combining the assumption of local realism with the quantum predictions (31d), (31c), and (31b) allows one to conclude that there must be a nonzero probability (with value at least as large as the value of $p13$) to obtain the results $a_1=-1$ and $b_1=-1$ in a joint measurement of $a_1$ and $b_1$. The remaining prediction in Eq.\ (31a), however, tells us that we cannot get simultaneously the results $a_1=-1$ and $b_1=-1$ in any run of the experiment. Hence a contradiction between quantum mechanics and local realism arises without using inequalities. A quite similar nonlocality argument to the one just described for the set of probabilities $\{p13,p4,p5,p9\}$ could be established for any one of the 64 sets $\{p\text{j},p\text{k},p\text{l},p\text{m}\}$ in Eqs.\ (27)-(30), provided we make $p\text{k}=p\text{l}=p\text{m}=0$ and $p\text{j}>0$. Further, it can be shown that, for \textit{any\/} of such sets, the maximum value of $p\text{j}$ predicted by quantum mechanics when $p\text{k}=p\text{l}=p\text{m}=0$ is $\tau^{-5}$, with $\tau$ being the golden mean, $(1+\sqrt{5})/2$. This gives an upper limit of $p\text{j}=0.090\,\, 17$.

It is worth noting that the ``upper'' inequalities $p{\text{k}}^{\prime}+p{\text{l}}^{\prime}+p{\text{m}}^{\prime}\leq 1+p{\text{j}}^{\prime}$ also play a role in revealing Hardy's nonlocality since a violation of the inequality $p\text{j} \leq p\text{k} +p\text{l} +p\text{m}$ for the case in which $p\text{k}=p\text{l}=p\text{m}=0$ and $p\text{j}>0$ necessarily implies a violation of the inequality $p{\text{k}}^{\prime}+p{\text{l}}^{\prime}+p{\text{m}}^{\prime}\leq 1+p{\text{j}}^{\prime}$ for some appropriate ${\text{k}}^{\prime}$, ${\text{l}}^{\prime}$, ${\text{m}}^{\prime}$, and ${\text{j}}^{\prime}$, the degree of this violation being such that the sum $p{\text{k}}^{\prime}+p{\text{l}}^{\prime}+p{\text{m}}^{\prime}$ on the left-hand side exceeds the sum $1+p{\text{j}}^{\prime}$ on the right-hand side by a quantity which is just equal to $p\text{j}$. To illustrate this fact we may consider the following numerically simple but significant example of values predicted by quantum mechanics for the various probabilities $p1,p2,\ldots,p16$ (see Appendix A of [5])
\begin{align}
p1& = 0.25,  & \, p2& = 0.375,  & \, p3& = 0.375, & \, p4& = 0, \,\,\,\,\,\, \nonumber  \\
p5& = 0,  & \, p6& = 0.625,  & \, p7& = 0.225, & \, p8& = 0.15, \,\,\,\,\,\,  \nonumber  \\[-.3cm]
& \, & \, & \, &  \, \\[-.3cm]
p9& = 0,  & \, p10& = 0.225, & \, p11& = 0.625, & \, p12& = 0.15, \,\,\,\,\,\, \nonumber  \\
p13& = 0.09,  & \, p14& = 0.135, & \, p15& = 0.135, & \, p16& = 0.64. \,\,\,\,\,\, \nonumber
\end{align}
Note that $p4=p5=p9=0$ and $p13=0.09$, so that the lower bound of the thirteenth inequality in (27), $p13\leq p4+p5+p9$, is violated. One might ask how many of the inequalities $p\text{j} \leq p\text{k} +p\text{l} +p\text{m} \leq 1+p\text{j}$ are violated by the above quantum mechanical values. It is easily verified that, for such values, all the 16 inequalities in Eq.\ (27) are violated (either from below or from above), whereas all the remaining 48 inequalities in Eqs.\ (28)-(30) are not. For eight of the inequalities (27) the violation entailed by the values (32) is of the form $p\text{i} \leq p\text{i} -0.09$ (specifically, for $\text{i}=2,3,6,7,10,11,13,16$), whereas for the remaining eight inequalities it is of the form $1+p{\text{i}}^{\prime}+0.09\leq 1+p{\text{i}}^{\prime}$ (${\text{i}}^{\prime}=1,4,5,8,9,12,14,15$). Also, the above values (32) make $\Sigma_1=0.82$ and $\Sigma_{1}^{\prime}=3.18$, and then both the CHSH inequalities $1\leq \Sigma_1$ and $\Sigma_{1}^{\prime}\leq 3$ are violated.

In general, if we have $p\text{k}=p\text{l}=p\text{m}=0$ and $0 < p\text{j}\leq 1/2$ \footnote{%%%%
The upper limit 1/2 comes from the fact that, according to the physical locality constraint of the kind (24), the probability $p\text{j}$ is restricted to lie in the range $[0,1/2]$ whenever the probabilities $p\text{k}$, $p\text{l}$, and $p\text{m}$ are equal to zero. Note, however, that this limit is considerably larger than that allowed by quantum mechanics, $0.090\,\, 17$.}
then each one of the 16 inequalities in the single set (27), (28), (29), or (30) containing the inequality $p\text{j} \leq p\text{k} +p\text{l} +p\text{m} \leq 1+p\text{j}$ in question, is automatically violated whereas the remaining 48 inequalities are not. In any case the violation acquires the form $p\text{i} \leq p\text{i} - p\text{j}$ (for eight of the inequalities in the set) or else $1+p{\text{i}}^{\prime}+ p\text{j}\leq 1+p{\text{i}}^{\prime}$ (for the remaining eight), with $\{p\text{i}\}\cup \{p{\text{i}}^{\prime}\} = \{p1,p2,\ldots,p16\}$. Furthermore, from relations (12)-(19), it follows that one of the $\Sigma_{i}$'s must be either $\Sigma_{i} = 1-2p\text{j}$ or $\Sigma_{i} = 3+2p\text{j}$ (correspondingly, $\Sigma_{i}^{\prime}$ must be either $\Sigma_{i}^{\prime}=3+2p\text{j}$ or $\Sigma_{i}^{\prime}=1-2p\text{j}$), and thus the CHSH inequality $1 \leq \Sigma_i, \Sigma_{i}^{\prime} \leq 3$ is violated for the appropriate index $i=1,2,3,\text{ or }4$. (For the above example (32) the index was $i=1$.) Conversely, from relations (12)-(19), it can be seen that a violation of the inequality $p{\text{k}}^{\prime}+p{\text{l}}^{\prime}+p{\text{m}}^{\prime}\leq 1+p{\text{j}}^{\prime}$ such that the left-hand side exceeds the right-hand side by $p\text{j}$ ($0<p\text{j}\leq 1/2$) causes three of the probabilities $p\text{k}$, $p\text{l}$, and $p\text{m}$ to be equal to zero, for some appropriate $\text{j}$, $\text{k}$, $\text{l}$, and $\text{m}$, so that the Bell-type inequality $p\text{j} \leq p\text{k} +p\text{l} +p\text{m}$ is correspondingly violated.

Let us finally mention that for the singlet, maximally entangled state quantum mechanics predicts $p\text{j}=0$ whenever we have $p\text{k}=p\text{l}=p\text{m}=0$, and then, as is well known, Hardy's argument does not work for this case. Equivalently, it can be shown that, for the singlet state, the fulfillment of the conditions $p\text{k}=p\text{l}=p\text{m}=0$ necessarily implies the perfect correlations $c(a_1,b_1)=\alpha$, $c(a_1,b_2)=\beta$, $c(a_2,b_1)=\gamma$, and $c(a_2,b_2)=\delta$, where $\alpha$, $\beta$, $\gamma$, and $\delta$ are either $+1$ or $-1$. These perfect correlations are such that the CHSH inequality, $-2\leq \Delta_i \leq 2$, is always satisfied for the singlet state whenever $p\text{k}=p\text{l}=p\text{m}=0$ [8].

\section{Equivalence of the CH and CHSH inequalities provided physical locality holds}

The CH inequalities are of the form [11-13]
\begin{equation}
-1 \leq B_i \leq 0,
\end{equation}
where $B_i$ can be any one of the four sums of probabilities
\begin{align}
& B_1 =p(a_1,b_1)+p(a_1,b_2)+p(a_2,b_1)-p(a_2,b_2)-p(a_1)-p(b_1),  \nonumber  \\
& B_2 =p(a_1,b_1)+p(a_1,b_2)-p(a_2,b_1)+p(a_2,b_2)-p(a_1)-p(b_2),   \nonumber \\[-.3cm]
& \\[-.3cm]
& B_3 =p(a_1,b_1)-p(a_1,b_2)+p(a_2,b_1)+p(a_2,b_2)-p(a_2)-p(b_1),  \nonumber  \\
& B_4 =-p(a_1,b_1)+p(a_1,b_2)+p(a_2,b_1)+p(a_2,b_2)-p(a_2)-p(b_2).  \nonumber  
\end{align}
In Eq.\ (34), $a_j$ and $b_k$, $j,k = 1,2$, denote the two possible settings for the two measuring apparatuses, and $p(a_j,b_k)$ denotes the joint probability that both particle 1 and particle 2 cross their respective analyzers with settings $a_j$ and $b_k$ and both are detected. $p(a_j)$ and $p(b_k)$ are the \textit{single\/} detection probabilities for particles 1 and 2, respectively, when the first (second) apparatus is set at $a_j$ ($b_k$). The analyzers can either transmit or absorb the incoming particles. This dichotomic choice can be used to define corresponding observables $a_j$ in such a way that the outcome $a_j=+1$ ($a_j=-1$) corresponds to transmission (absorption) of particle 1 when it interacts with the analyzer with setting $a_j$. Observables $b_k$ are similarly defined for particle 2. Thus, for the ideal experimental conditions we are considering, the following identifications can be made
\begin{align}
\,\,p(a_1,b_1) &= p(a_1=1,b_1=1)\equiv p1, & \,\,\, p(a_1,b_2) &= p(a_1=1,b_2=1)\equiv p5, \,\,\,\,\,\, 
\,\,\,\,\,\,\, \nonumber  \\[-.3cm]
& \\[-.3cm]
\,\,p(a_2,b_1) &= p(a_2=1,b_1=1)\equiv p9, & \,\,\, p(a_2,b_2) &= p(a_2=1,b_2=1)\equiv p13. \,\,\,\,\,\,
\,\,\,\,\,\,\, \nonumber
\end{align}
Furthermore, the single probabilities $p(a_j)$ and $p(b_k)$ can be substituted by (see Eqs.\ (10a)-(10b))
\begin{align}
p(a_j)&= p(a_j=1,b_1=1)+p(a_j=1,b_1=-1),  \tag{36a}  \\
&= p(a_j=1,b_2=1)+p(a_j=1,b_2=-1),  \tag{36b}
\setcounter{equation}{36}
\end{align}
and 
\begin{align}
p(b_k) &= p(a_1=1,b_k=1)+p(a_1=-1,b_k=1),  \tag{37a}  \\
       &= p(a_2=1,b_k=1)+p(a_2=-1,b_k=1).  \tag{37b}
\setcounter{equation}{37}
\end{align}
Now, by using Eqs.\ (35)-(37) and recalling (3), we may rewrite the $B_i$'s in Eq.\ (34) in terms of the joint probabilities $p1,p2,\ldots,p16$. Since we may use either one of equations (36a) or (36b) [(37a) or (37b)] for $p(a_j)$ [$p(b_k)$], there will be four equivalent forms of expressing each $B_i$. Next we write down one of the possible forms for the $B_i$'s, each involving \textit{four} joint probabilities
\begin{align}
& B_1 =-p2+p5-p11-p13,  \nonumber  \\
& B_2 =-p2+p5-p9-p15,  \nonumber  \\[-.3cm]
& \\[-.3cm]
& B_3 =-p3-p5-p10+p13,  \nonumber  \\
& B_4 =-p1+p5-p10-p15.  \nonumber  
\end{align}
Then, from Eq.\ (38), and making use of the appropriate physical-locality-induced relation in Eqs.\ (12)-(19), one finds the following general expression for the quantity $B_i$
\begin{equation}
B_i = \frac{1}{2} (\Sigma_i -3) = \frac{1}{2} (1-\Sigma_{i}^{\prime}), \quad i=1,2,3,4.
\end{equation}
Finally, by inserting Eq.\ (39) into the CH inequality (33), we immediately retrieve the CHSH inequality in Eq.\ (9), $1 \leq \Sigma_i, \Sigma_{i}^{\prime} \leq 3$. It is to be noted, incidentally, that, strictly speaking, we may equally arrive at expression (39) by confining ourselves to either one of the expressions (36a) or (36b) [(37a) or (37b)] for $p(a_j)$ [$p(b_k)$]. For convenience, however, we have made use of the property of physical locality from the outset in order to achieve the simplest expression for each of the $B_i$'s, namely one involving only four joint probabilities. In any case, one must necessarily rely on this property when passing from Eq.\ (38) to Eq.\ (39). 

Conversely, it can readily be shown that, granted physical locality, the CHSH inequality (9) implies the CH inequality (33). Clearly, to do this, it suffices to see that $\Sigma_i = 2B_i +3$ (cf.\ Eq.\ (39)). Let us consider for concreteness the case of $\Sigma_1$. So, from the definition in (4) and the normalization condition (2) we may write $\Sigma_1$ as $3-p2-p3+p5+p8-p10-p11-p13-p16$. Now, from the easily confirmed relationship entailed by physical locality (see Eq.\ (13)), $p8-p3-p10-p16 = p5-p2-p11-p13$, and the above expression for $B_1$ in (38), we find that $\Sigma_1 = 2B_1 +3$.

As a corollary, we mention that the Bell-type inequalities $p\text{j} \leq p\text{k} +p\text{l} +p\text{m} \leq 1+p\text{j}$ in Eqs.\ (27)-(30) are completely equivalent to both the CH and CHSH inequalities, provided physical locality holds. As a concrete example of this equivalence let us note that, from the expression of $B_2$ in (38) and the CH inequality $-1\leq B_2 \leq 0$, we readily obtain, $p5\leq p2+p9+p15\leq 1+p5$. Conversely, the inequality $p5\leq p2+p9+p15\leq 1+p5$ can also be written as $-1\leq -p2+p5-p9-p15\leq 0$,\linebreak thus obtaining $-1\leq B_2 \leq 0$. On the other hand, from the CHSH inequality $1\leq \Sigma_2 \leq 3$ and the physical locality relation $\Sigma_2 =1+2(p2-p5+p9+p15)$, we obtain the inequality $p5\leq p2+p9+p15\leq 1+p5$. Conversely, from this latter inequality and the relation $p2-p5+p9+p15=(\Sigma_2 -1)/2$ we get immediately the CHSH inequality $1\leq \Sigma_2 \leq 3$.

\section{The question of Popescu and Rohrlich revisited}

Popescu and Rohrlich [17] addressed the question of whether relativistic causality restricts the maximum quantum prediction of the CHSH sum of correlations to $2\sqrt2$ instead of 4. As Popescu and Rohrlich put it [17], ``Rather than ask why quantum correlations violate the CHSH inequality, we might ask why they do not violate it \textit{more}.'' They found that relativistic causality does not by itself constrain the maximum CHSH sum of quantum correlations to $2\sqrt2$, and gave a set of joint probabilities which satisfies the causal communication constraint and which provides the maximum level of violation, 4. Their set of probabilities is that in Eq.\ (20). For this set we have
\begin{equation}
\Sigma_1 =p1+p4+p5+p8+p9+p12+p14+p15=4,
\end{equation}
and
\begin{equation}
B_1 = -p2+p5-p11-p13 = 1/2.
\end{equation}
The set of values $p5=1/2$ and $p2=p11=p13=0$ satisfies, as it should be, the inequality of the form (24) required by physical locality
\begin{equation}
2\,p5 -1 \leq p2+p11+p13  .
\end{equation}
However, as pointed out by Mermin in discussing the question of Popescu and Rohrlich [10], such probabilities do not yield a maximal violation of the CH-type inequality \footnote{%%%
Following Mermin [10], in this section the Bell-type inequality $p\text{j} \leq p\text{k} +p\text{l} +p\text{m}$ will be referred to as being of the ``CH-type'' in spite of the fact that, as we have seen, it can equally be regarded as being of the ``CHSH-type'', provided physical locality holds.}
\begin{equation}
p5 \leq p2+p11+p13 ,
\end{equation}
since the resulting 50\% violation is less than the 100\% maximal violation entailed by the values $p5=1$ and $p2=p11=p13=0$. By contrast, for the case of the CHSH inequality
\begin{equation}
p1+p4+p5+p8+p9+p12+p14+p15 \leq 3,
\end{equation}
the physically local pair distribution in (20) does indeed make the sum on the left-hand side to be equal to its upper probabilistic limit, namely 4 (cf.\ Eq.\ (23) and (40)). In view of the equivalence of the CH and CHSH inequalities, the set of probabilities (20) deduced by Popescu and Rohrlich might seem somehow pathological in that they give the maximum level of violation of the CHSH inequality (44) while they are, seemingly, unable to violate maximally the equivalent CH inequality (43). As properly remarked in [10], however, there is nothing paradoxical in this situation because, as shown explicitly in Sec.\ 5, the proof of equivalence between the inequalities (43) and (44) rests heavily on the property of physical locality. Therefore, the distributions of probabilities for which the set of values $\{p5,p2,p11,p13\}$ violate the physical locality constraint in Eq.\ (42), do not correspond to any permissible combination of correlations and thus should not appear in either the CH or CHSH inequalities. Such unphysical, ultra-nonlocal distributions include, in particular, those for which $p2=p11=p13=0$ \textit{and\/} \mbox{$1/2< p5 \le 1$}. It is to be emphasized, however, that, as can easily be seen from the expression of $p5$ in Eq.\ (13), the local pair distribution of probabilities for which $p2=p11=p13=0$ and $p5=1/2$ does indeed imply the values $p1=p4=p5=p8=p9=p12=p14=p15=1/2$, and then such a distribution gives the maximum level of violation of the CHSH inequality (44). Conversely, the local pair distribution for which $p1=p4=p5=p8=p9=p12=p14=p15=1/2$ does equally imply the maximum permissible level of violation of the CH inequality (43) since now we have $B_1 =p5-p2-p11-p13 = (\Sigma_1 -3)/2 = 1/2$.

As noted by Kwiat and Hardy [26], there is a fair amount of freedom in constructing ultra-nonlocal pair distributions which are subjected to the sole conditions $p\text{k}=p\text{l}=p\text{m}=0$ (of course, it is also understood that $1/2 < p\text{j} \leq 1$), the set of probabilities $\{p\text{j},p\text{k},p\text{l},p\text{m}\}$ being any one of the 64 combinations of four probabilities entering the Bell-type inequalitites $p\text{j} \leq p\text{k} +p\text{l} +p\text{m}$ in Eqs.\ (27)-(30). One rather peculiar example of such an ultra-nonlocal distribution for which $p2=p11=p13=0$ and $p5=1$ is [26]
\begin{equation}
p1=p5=p9=p14=1,
\end{equation}
and all other probabilities zero. For the values (45) we have $\Sigma_1 =4$. On the other hand, from the definition of $B_1$ in (34) we find $B_1 =1$. It is important to notice that, although the expression in Eq.\ (41) gives, coincidentally, the value $B_1 =1$ for the distribution (45), the formula (41) for $B_1$ cannot be applied to the case of non-local pair distributions because such a formula is based on the assumption of physical locality. In fact, for the values (45), $B_1 \neq (\Sigma_1 -3)/2$ so that the CH and CHSH inequalities in Eqs.\ (43) and (44) are no longer equivalent. It is worth noting, incidentally, that one could actually exploit the ultra-nonlocal correlations in (45)
\begin{equation}
p5\equiv p(a_1=1, b_2=1)=1   \quad\text{and}\quad   p14\equiv p(a_2=1,b_2=-1)=1,
\end{equation}
to convey superluminal signals. This follows from the fact that, according to (46), the result obtained in a measurement of the observable $b_2$ at wing $B$ is completely determined by the type of measurement performed at the spacelike separated wing $A$. In general, however, as was noted in Sec.\ 3, the ultra-nonlocal correlations require one to test a statistically significant number of particles at one side in order to ascertain the observable being measured at the other.

\section{Conclusion}
We conclude by recalling the fact that, as noted in Sec.\ 4, there are four inequalities of the form $p\text{j} \leq p\text{k} +p\text{l} +p\text{m}$ for each $\text{j}=1,2,\ldots,16$. This means that, for example, the fulfillment of either one of the two sets of conditions (see the first inequality in (27) and (28))
\begin{align}
p6 &\equiv p(a_1=+1, b_2=-1) =0,  \nonumber  \\
p11 &\equiv p(a_2=-1, b_1=+1)=0,   \nonumber \\[-.3cm]
&  \\[-.3cm]
p13 &\equiv p(a_2=+1, b_2=+1)=0,   \nonumber \\
p1 &\equiv p(a_1=+1, b_1=+1)>0,  \nonumber
\end{align}
or
\begin{align}
p6 &\equiv p(a_1=+1, b_2=-1) =0,  \nonumber  \\
p9 &\equiv p(a_2=+1, b_1=+1)=0,   \nonumber \\[-.3cm]
&  \\[-.3cm]
p15 &\equiv p(a_2=-1, b_2=+1)=0,   \nonumber \\
p1 &\equiv p(a_1=+1, b_1=+1)>0,  \nonumber
\end{align}
leads to a nonlocality contradiction of the Hardy type. We can see that the sets of conditions (47) and (48) are ``almost'' identical, the only difference between them being the sign of the outcome for the observable $a_2$. One could certainly find out many more instances of this kind of similarity existing between the sets of probabilities $\{p\text{j},p\text{k},p\text{l},p\text{m}\}$ in Eqs.\ (27)-(30). This shows that the class of correlations which can exhibit Hardy's nonlocality is actually much subtler and richer than one might have thought.

In summary, in this paper we have derived systematically all the Bell-type inequalities (each involving four joint probabilities) which can be used to experimentally test Hardy's nonlocality for two photons or particles with spin 1/2. To do this, we have made extensive use of the experimentally testable property of physical locality through the constraints it imposes on the the joint probabilities involved in the EPR-Bohm experiment, Eqs.\ (12)-(19). It is to be emphasized that all these constraints (as well as all those in Eqs.\ (21)-(24)) are rather general, and should be fulfilled by any physical theory consistent with relativistic causality (the nonnegativity of the probabilities taken for granted). In fact, a violation of the requirement of physical locality would entail the existence of ultra-nonlocal correlations which could be used to set up superluminal communication. \enlargethispage{-.3cm}Further, we have provided a detailed proof of the fact that the Bell inequalities for the Hardy experiment, Eqs.\ (27)-(30), are completely equivalent to both the CH and CHSH inequalities, provided physical locality holds. This state of affairs, however, changes drastically in the absence of physical locality: The CH and CHSH inequalities are no longer equivalent, the relations in Eqs.\ (12)-(19) are no longer valid, and hence the inequalities in Eqs.\ (27)-(30) do not follow from either the CH or CHSH inequalities any more.\,\footnote{%%%
I hope, therefore, that the principle of physical locality holds true at least until the refereeing of this paper is completed!},\,\footnote{%%%
More seriously, it should be added that neither the CH inequality nor the CHSH inequality themselves can be established in the event that the physical locality property does not follow. This is so because, as first shown by Jarrett [27], the factorizability condition (or \textit{strong locality}) on which the derivation of such inequalities is based is logically equivalent to the conjunction of the conditions of ``locality'' and ``completeness'', the condition of locality being just the condition of physical locality expressed in Eqs.\ (10a)-(10b). Incidentally, I prefer Shimony's terminology [28] of ``parameter independence'' and ``outcome independence'' to refer to Jarrett's conditions of ``locality'' and ``completeness''.}

\vspace{.2cm}
\textbf{Acknowledgments} --- The author wishes to thank the anonymous referees for their valuable suggestions which led to an improvement of an earlier version of this paper.

\end{document}